\documentstyle[12pt]{article}

\def\bseq{\begin{subequation}}  % = 1a 1b
\def\eseq{\end{subequation}}
\def\bsea{\begin{subeqnarray}}  % = 1.1a 1.1b
\def\esea{\end{subeqnarray}}

\def\beq{\begin{equation}}
\def\eeq{\end{equation}}
\def\eea{\end{eqnarray}}
\def\bq{\begin{quote}}
\def\eq{\end{quote}}

\newcommand{\EQ}{\begin{equation}}
\newcommand{\EN}{\end{equation}}
\newcommand{\bea}{\begin{eqnarray}}
\newcommand{\ena}{\end{eqnarray}}

\renewcommand{\a}{\alpha}
\renewcommand{\b}{\beta}

\renewcommand{\d}{\delta}
\newcommand{\th}{\theta}

\newcommand{\pa}{\partial}

\newcommand{\e}{\epsilon}

\newcommand{\r}{\rho}

\newcommand{\s}{\sigma}

\newcommand{\reff}[1]{eq.(\ref{#1})}

\newcommand{\shalf}{\frac{1}{2}}
\newcommand{\bl}[1]{\Bigl(#1\Bigr)}
\newcommand{\blt}[1]{\Bigl( #1\Bigr)_\newh}
\newcommand{\cbl}[1]{\Bigl\{ #1\Bigr\}_\newh}
\def\newh{H}
\def\mt{\widetilde{m}}

\def\sh{\mbox{sh}}
\def\ch{\mbox{ch}}

\newcommand{\NP}[1]{Nucl.\ Phys.\ {\bf #1}}
\newcommand{\PL}[1]{Phys.\ Lett.\ {\bf #1}}

\renewcommand{\thefootnote}{\fnsymbol{footnote}}

\begin{document}
\newpage
\begin{titlepage}
\begin{flushright}
{CERN-TH.6333/91}\\
{IFUM 409/FT}
\end{flushright}
\vspace{2cm}
\begin{center}
{\bf {\large EXACT S-MATRICES FOR THE NONSIMPLY-LACED}} \\
\vspace{.1in}
{\bf {\large AFFINE TODA THEORIES $a^{(2)}_{2n-1}$ }} \\
\vspace{1.5cm}
{G.W. DELIUS and
M.T. GRISARU}\footnote{On leave from Brandeis University, Waltham, MA 02254,
USA\\Work partially supported by the National Science Foundation under
grant PHY-88-18853.}\\
\vspace{1mm}
{\em Theory Division, CERN, 1211 Geneva 23, Switzerland}\\

\vspace{5mm}
and

\vspace{5mm}
{D. ZANON} \\
\vspace{1mm}{\em Dipartimento di Fisica dell' Universit\`{a} di Milano and} \\
{\em INFN, Sezione di Milano, I-20133 Milano,
Italy}\\
\vspace{1.1cm}
{\bf{ABSTRACT}}
\end{center}
\bq
We derive the exact, factorized, purely elastic scattering matrices for the
$a^{(2)}_{2n-1}$ family of nonsimply-laced affine Toda theories.
The derivation takes into account the distortion of the classical mass
spectrum by radiative corrections, as well as modifications of the usual
bootstrap assumptions since for these theories anomalous threshold
singularities
lead to a displacement of some  single particle
poles.
\eq

\vfill
\begin{flushleft}
CERN-TH.6333/91\\
IFUM 409/TH\\
November 1991
\end{flushleft}
\end{titlepage}

\renewcommand{\thefootnote}{\arabic{footnote}}
\setcounter{footnote}{0}
\newpage

Affine Today theories are massive two-dimensional field theories
represented by lagrangians of
the form
\EQ
{\cal L}= -\frac{1}{2\b^2} \vec{\phi} \Box \vec{\phi} -\sum \frac{q_i}{\b^2}
 e^{\vec{\a}_i\cdot
 \vec{\phi}}
 \EN
Here  the
$\vec{\a}_i$ are the simple roots of a rank $r$ Lie algebra augmented by
(the negative of) a maximal root and $\vec{\phi}=(\phi_1,\phi_2,...\phi_r)$
are bosonic fields
describing $r$ massive particles, with classical masses rigidly controlled by
properties of the Lie algebra. The constants $q_i$ are such that
$\sum q_i \vec{\a}_i=0$ and $\b$ is the coupling constant.
These theories are classically
integrable: the lagrangian above admits an infinite number of symmetries
described by conserved currents $J_{\pm}^{(s)}$ of increasing spin $s$.
If at least one of these  symmetries with $s>2$ survives quantization,
the $n$-particle  scattering matrices of these theories
factorize into a product
of elastic  two-particle S-matrices.
Assumptions of analyticity, unitarity and a
bootstrap principle \cite{1} should allow the full S-matrix to be
determined exactly. Indeed, exact
S-matrices have been proposed for all affine Toda theories based on
simply-laced Lie algebras and perturbative calculations based on the
corresponding lagrangians have verified the correctness of these
proposals \cite{2,3,3p}.
On the other hand arguments have been put forward to suggest that a
corresponding construction would fail for affine Toda theories based on
nonsimply-laced algebras and indeed, so far, no satisfactory
S-matrices have been found for these theories.

There are several reasons for the belief  that nonsimply-laced Toda theories
might not have exact factorizable S-matrices: a) Usually the  bootstrap
procedure leads to restrictions on the mass spectrum of the particles.
These restrictions are consistent with the classical masses, and indeed for
simply-laced theories the mass spectrum does not seem to be
affected by radiative corrections (except for an overall, irrelevant
rescaling). On the other hand, for the nonsimply-laced theories
radiative corrections distort the mass spectrum in a manner that at
first sight seems incompatible with bootstrap restrictions \cite{2,3,4}.
b) Exact
S-matrices contain higher-order poles which must be explained as
anomalous threshold singularities of corresponding Feynman diagrams
(or, in a pure S-matrix approach, of extended unitarity-analyticity
diagrams \cite{4p}). However, for the nonsimply-laced theories such an
explanation has failed to account for all the singularities of proposed
S-matrices \cite{2}. c) We have recently  pointed out that certain anomalous
threshold singularities may lead to a breakdown of some of the bootstrap
assumptions \cite{5}.

In view of these facts it may be reasonable to believe that quantum
effects destroy the classical integrability of these theories: the
classical conservation laws which imply the factorization and elasticity
of the scattering matrix do not survive quantization. However this is
not the case.
A study
of the renormalization of the higher spin currents has revealed  that
for example in the
$a^{(2)}_3$ nonsimply-laced Toda theory a non-anomalous higher
spin symmetry does
exist \cite{6} and this implies that at least in this case it should be
possible to construct a factorized S-matrix.

In fact, the problems mentioned above concerning the mass
spectrum, anomalous threshold structure and breakdown of the bootstrap
provide each other's solution.
We have been able  to construct an
exact factorized S-matrix for the $a^{(2)}_{2n-1}$ family of nonsimply-laced
Toda theories and we outline this construction here; further details, the
perturbative verification, and the construction of quantum-conserved currents
will be presented elsewhere \cite{6,7}. For general information about the
techniques used the reader is referred to the considerable literature that has
built up in recent years; for a review and references see \cite{8}.

The Toda theory
based on the nonsimply-laced twisted affine Lie algebras $a_{2n-1}^{(2)}$
is described by the lagrangian
\bea\label{action}
\b^2 {\cal{L}} &=& - {\frac12}\sum_1^{n-1} \phi_a
\Box \phi_a -{\frac12} \phi_n \Box \phi_n\\
&-& \sum_{k=2}^{n-1} 2 \exp{\left(-\sqrt{\frac{2}{h}} \sum_{1}^{n-1}
m_a \cos{\frac{(2k-1)a \pi}{h}} ~\phi_a \right)}
- \exp{\left( -\sqrt{\frac{2}{h}} \sum_{1}^{n-1} (-1)^a m_a \phi_a
\right)}\nonumber\\
&-& \exp{\left( -\sqrt{\frac{2}{h}} \sum_{1}^{n-1} m_a \cos{\frac{a \pi}
{h}} ~\phi_a + \phi_n \right)}
-  \exp{\left( -\sqrt{\frac{2}{h}} \sum_{1}^{n-1} m_a \cos{\frac{a\pi}
{h}} ~\phi_a - \phi_n \right)}\nonumber
\ena
with masses and
three-point couplings given by
\beq\label{m}
m_a^2 = 8\sin^2{(\frac{a\pi }{h})}  ~~~~~~~a=1,...,n-1 ,\qquad
m_n^2 = 2
\eeq
\bea\label{c}
c_{abc} &=& -\frac{1}{\sqrt{2h}}m_a m_b m_c ~~~~~~~~~~~~{\rm{if}} ~~a+b+c=2n-1
\nonumber \\
c_{abc} &=& \frac{1}{\sqrt{2h}}m_a m_b m_c ~~~~~~~~~~~~~~{\rm{if}} ~~a\pm b
\pm c=0
\nonumber \\
c_{ann} &=& \sqrt{\frac{2}{h}} m_n^2 m_a \cos{\frac{a \pi}{h}}
\ena
All other three-point couplings are zero.
We have introduced the Coxeter number $h=2n-1$ and set the overall mass-scale
to $1$.

At the one-loop level radiative corrections lead to mass shifts that
have been computed in Ref. \cite{4}. We absorb the shift in $m_n$
into an overall mass rescaling, in which case the results in eqs.(20,22)
of the above reference become
\beq\label{dm}
\d  m_a^2= -\frac{2a}{h^2}\sin \frac{2a\pi}{h} ,\qquad
\d m_n^2 =0
\eeq
In contrast,  for a simply-laced theory
the mass corrections   vanish after the above rescaling.

For the specific case $n=2$, i.e. the $a_3^{(2)}$ theory, the lagrangian
is
\EQ
\b^2{\cal L} =-{\frac12} \phi_1\Box\phi_1 -\\{\frac12}\phi_2\Box\phi_2 -
e^{-\phi_1-\phi_2} - e^{-\phi_1+\phi_2} -e^{2\phi_1}
\EN
Using the classical field equations it is straightforward to check
the conservation
$\pa_-J_+^{(s)} +\pa_+J_-^{(s)}=0$
of spin 2 and spin 4 currents, where the spin 2 current is the
stress tensor, while
\EQ
J_+^{(4)}\equiv J_{++++}=(\pa_+\phi_1)^2(\pa_+\phi_2)^2 +(\pa_+^2\phi_2)^2
+2\pa_+\phi_1\pa_+\phi_2\pa_+^2\phi_2
\EN
with a suitable expression for $J_-^{(4)}$. At the quantum level we have
shown \cite{6} that the
conservation laws still hold with a renormalized current
\bea\label{rJ}
J_+^{(4)} &=& (1+\frac{\hbar}{2})(\pa_+\phi_1)^2(\pa_+\phi_2)^2
-\frac{\hbar}{12}(1+3\hbar+\hbar^2)(\pa_+^2\phi_1)^2
-\frac{\hbar}{12} (\pa_+\phi_1)^4 -\frac{\hbar}{12} (\pa_+\phi_2)^4 \nonumber\\
&&+(1+\frac{23}{12}\hbar +\hbar^2 +\frac{1}{6}\hbar^3)
(\pa_+^2\phi_2)^2+(2+3\hbar +\hbar^2)
\pa_+\phi_1\pa_+\phi_2\pa_+^2\phi_2
\ena
We expect that a similar conserved current exists for all the
$a_{2n-1}^{(2)}$ Toda theories.
Together with the renormalized stress tensor this defines corresponding
spin 1 and spin 3 charges
%\EQ
%Q^{(1)}=\int dx^+T_{++} ~~~,~~~Q^{(3)}=\int dx^+ J_{++++}
%\EN
whose presence should  guarantee the existence of factorizable
elastic S-matrices. The bootstrap procedure should then determine them.
However, there are some subtleties \cite{5}:

The current conservation implies that in a scattering process the sum
of the charges of incoming particles equals that of the outgoing
particles. The bootstrap principle, which asserts that in the
two-body S-matrix a simple pole is to be associated with a particle in
the spectrum via the process $a+b \rightarrow c \rightarrow a+b$
extends this to the vertex
function $<a,b,c>$ with three on-shell particles.
In particular, in the $n=2$ theory described above, the existence of the
conserved spin 4 current would imply that either the spin 3 charge of particle
$\phi_1$ is zero, or else the corresponding vertex function $<1,1,1>$ vanishes.
An examination of the currents above reveals that at the classical level the
charge does indeed vanish, which is consistent with the presence of a
$\phi_1^3$ coupling, but loop calculations reveal an apparent
inconsistency: the charge is not zero \cite{5,6}, and the vertex function does
not vanish. These  results can be explained in fact by the presence of an
anomalous threshold singularity in a triangle graph which renders the vertex
function infinite on shell, but they do indicate that one has to be
careful in applying bootstrap ideas to processes where the $<a,b,h-a-b>$ vertex
enters.

We shall construct the S-matrix by following procedures similar to those
used in the simply-laced case, but with two important differences: we
will admit that it has simple particle poles at positions shifted away
from the classical mass values, and we shall relax the bootstrap
principle since some simple poles are shifted
away from their single-particle positions due to anomalous threshold
effects. We will have to prove however that all the singularities we
find can be accounted for in this fashion.

Assuming that higher-spin quantum-conserved currents similar to those in
\reff{rJ} exist for all the $a_{2n-1}^{(2)}$ theories, we postulate
the existence of purely elastic two-body amplitudes.
Imposing
unitarity and real analyticity on the S-matrix for the process $a+b
\rightarrow a+b$ restricts
$S_{ab}(\th)$ to be a product of fundamental building blocks
\beq\label{blocks}
S_{ab}(\th)=\prod_{x\in A_{ab}}\bl{x}
\qquad\qquad\mbox{where}\qquad
\bl{x}\equiv{\sh\left({\th\over2}+{i\pi\over2h}x\right)
\over\sh\left({\th\over2}-{i\pi\over2h}x\right)}
\eeq
and $\th=\th_a-\th_b$ is the relative rapidity.
We are using the notation of reference \cite{2}.
This notation is related
to that of reference \cite{3,3p} by $f_\a\equiv\bl{h\a}$. Crossing symmetry
acts on the blocks by $\bl{x}\rightarrow -\bl{h-x}$.

We start by determining the element $S_{nn}$.
The three-point couplings in \reff{c} suggest that all the particles
$\phi_1\cdots\phi_{n-1}$ appear as intermediate particles in this process,
both in
the direct and in the crossed channel. They lead to poles
$\left((p+q)^2-\widetilde{m}_a^2\right)^{-1}$ and
$\left((p-q)^2-\widetilde{m}_a^2\right)^{-1}$ in $S_{nn}$. Here
$\widetilde{m}_a$ are
the radiatively corrected masses. Without loss of generality we write, with
reference to \reff{m}
\beq\label{renmasses}
\widetilde{m}_a^2=4 \widetilde{m}_n^2\sin^2\left({\pi\over
h}\left(a+\e_a(\b)\right)\right)
\eeq
In the rapidity plane these poles are located at:
\bea\label{Snnpoles}
\mbox{s-channel:}\qquad 2\widetilde{m}_n^2(1+\ch\th)-\widetilde{m}_a^2=0
&&\Rightarrow\qquad\th={i\pi\over h}(h-2a-2\e_a)\nonumber\\
\mbox{u-channel:}\qquad 2\widetilde{m}_n^2(1-\ch\th)-\widetilde{m}_a^2=0
&&\Rightarrow\qquad\th={i\pi\over h}(2a+2\e_a)
\eea
We must reproduce these poles with the building blocks in
\reff{blocks}. Noting that $\bl{x}$ has a pole at $\th={i\pi\over
h}x$  we choose
%\beq\label{Snnmin}
%S_{nn}^{(min)}=\prod_{a=1}^{n-1}\ \bl{2a+2\e_a}\,\bl{h-2a-2\e_a}
%\eeq
%which has the right pole structure. However the S-matrix should reduce to the
%identity matrix when the coupling constant $\b$ is zero. This is achieved by
%the Ansatz
\beq\label{Snn}
S_{nn}=\prod_{a=0}^{n-1}\ {\bl{2a+2\e_a}\,\bl{h-2a-2\e_a}
\over\bl{2a+2\eta_a}\,\bl{h-2a-2\eta_a}}
\eeq
where $\e_a$ and $\eta_a$ are both zero when $\b=0$ so that the S-matrix
reduces to the identity matrix. This $S_{nn}$ is crossing symmetric. We let
the product start at $a=0$ for generality. None of the extra blocks which we
introduced in \reff{Snn} should produce any additional poles on the physical
sheet (i.e., for $0<\th<i\pi$) and this requires \beq\label{nopoles}
{h\over2}-a\geq\eta_a\geq
-a\qquad\mbox{and}\qquad \e_0\leq 0.
\eeq
Eventually, comparison to perturbation theory  will justify this
Ansatz.

We now turn to the determination of the S-matrix elements $S_{an},\ {a=1\cdots
n-1}$. These are  determined by $S_{nn}$ through the bootstrap
principle \cite{1}
\beq\label{bootSan}
S_{an}=S_{nn}(\th+{1\over2}\th_{nn}^a)S_{nn}(\th-{1\over2}\th_{nn}^a)
\eeq
where $\th_{nn}^a$ is the relative rapidity at which $S_{nn}$ has the s-channel
pole corresponding to particle $a$, i.e., $\th_{nn}^a={i\pi\over
h}(h-2a-2\e_a)$, see \reff{Snnpoles}.
%Using
%\beq
%\bl{x}(\th+{i\pi\over h}\r)\cdot\bl{x}(\th-{i\pi\over h}\r)=
%\bl{x+\r}(\th)\cdot\bl{x-\r}(\th).
%\eeq
We find
\bea\label{Sanlong}
S_{an}=\ \prod_{p=0}^{n-1}
&&{\bl{2p+2\e_p-{h\over2}+a+\e_a}\bl{h-2p-2\e_p-{h\over2}+a+\e_a}
\over\bl{2p+2\eta_p-{h\over2}+a+\e_a}\bl{h-2p-2\eta_p-{h\over2}+a+\e_a}}
\nonumber\\
\times&&{\bl{2p+2\e_p+{h\over2}-a-\e_a}\bl{h-2p-2\e_p+{h\over2}-a-\e_a}
\over\bl{2p+2\eta_p+{h\over2}-a-\e_a}\bl{h-2p-2\eta_p+{h\over2}-a-\e_a}}
\eea
This expression has a large number of poles, many more than perturbation theory
can be expected to explain. However for special values of $\e_a$ and $\eta_a$
many of the building blocks cancel each other.
We can not choose $\e_a=\eta_a$ because that would cancel the wanted poles in
\reff{Snn}. One allowed choice which cancels many poles in \reff{Sanlong} is
$\e_a=0$ and $\eta_a={B\over2}$ for all $a$ and some $B=B(\b)$.
%Then we are left
%with \beq
%S_{an}^{\rm unren}=\prod_{p=1}^{2a-1}\ {\bl{{h\over2}-a+p}^2\over
%\bl{{h\over2}-a+p+B}\,\bl{{h\over2}-a+p-B}}\
%{\bl{{h\over2}-a}\over\bl{{h\over2}-a+B}}\,
%{\bl{{h\over2}+a}\over\bl{{h\over2}+a-B}}
%\eeq
This choice corresponds to unrenormalized mass ratios. We note
however that there is a choice which has even fewer poles\footnote{Choosing
$\eta_0=0$ instead of $\eta_0=-h\e$ leads to the S-matrix for the
$A^{(4)}(0,2n)$ Toda theory, as we discuss at the end of the paper.}:
\beq\label{choice}
\e_a=a\e\ ,\qquad \eta_a=(a-h)\e\ ,\qquad a=0,\cdots,n-1
\eeq
$\e$ depends on $\b$ and satisfies $-\frac{1}{2n}\leq\e\leq 0$ in order to
fulfill \reff{nopoles}. This reduces \reff{Sanlong} to
\bea\label{San}
S_{an}=\prod_{p=1}^{a-1}&&{\bl{{h\over2}+(-a+2p)(1+\e)}^2\over
\bl{{h\over2}+(-a+2p)(1+\e)-2h\e}\,\bl{{h\over2}+(-a+2p)(1+\e)+2h\e}}
\nonumber\\
\times&&
{\bl{{h\over2}-a(1+\e)}\over\bl{{h\over2}-a(1+\e)-2h\e}}\,
{\bl{{h\over2}+a(1+\e)}\over\bl{{h\over2}+a(1+\e)+2h\e}}
\eea

This
expression still has two simple poles at $\th={i\pi\over
h}\left({h\over2}+a(1+\e)\right)$ and at $\th={i\pi\over
h}\left({h\over2}-a(1+\e)\right)$ as well as several double poles. We discuss
the double poles later on and note here that the single poles
correspond to particle $\phi_n$ in the intermediate s- and u-channels.
Indeed
\beq
s-\widetilde{m}_n^2=\widetilde{m}_n^2+\widetilde{m}_a^2+2\widetilde{m}_n
\widetilde{m}_a\ch\th
-\widetilde{m}_n^2=0\qquad\mbox{at }\ \th={i\pi\over
 h}\left({h\over2}+a(1+\e)\right)
\eeq
and similarly for the u-channel.

Let us introduce some simplifying notation:
\bea
\newh &=&{h\over 1+\e}\qquad ,\qquad B=-2h{\e\over 1+\e}\\
\blt{x}&=&{\sh\left({\th\over2}+{i\pi\over2\newh}x\right)
\over\sh\left({\th\over2}-{i\pi\over2\newh}x\right)}\qquad ,\qquad
\cbl{x}={\blt{x-1}\blt{x+1}\over\blt{x-1+B}\blt{x+1-B}}
\eea
and write \reff{Snn} and \reff{San} as
\bea\label{nSnn}
S_{nn}&=&\prod_{a=0}^{n-1}\ {\blt{2a}\,\blt{\newh-2a}
\over\blt{2a+B}\,\blt{\newh-2a-B}}\\
\label{nSan}
S_{an}&=&\prod_{p=1}^a\ \cbl{{\newh\over2}+2p-a-1}=S_{na}.
\eea
In terms of this notation \reff{renmasses}  implies that the renormalized
masses have the same form as the classical masses, but again
with $h$ replaced by $\newh$.

The remaining S-matrix elements $S_{ab},\ a,b=1\cdots n-1$ are obtained by
another application of the bootstrap
\bea\label{bootSab}
S_{ab}(\th)&=&S_{nb}\left(\th+\shalf\th_{nn}^a\right)
S_{nb}\left(\th-\shalf\th_{nn}^a\right)\nonumber\\
&&=\prod_{p=1}^b\ \cbl{2p-b-1+a}\,\cbl{\newh+2p-b-1-a}\label{nSab}
\eea
This expression is symmetric in $a,b$ as one can verify by using relations
such as $\prod_{p=1}^x\cbl{2p-x}=1$. Crossing symmetry can be
easily checked by a change of variable $p\rightarrow-p+b+1$ in one of the terms
in \reff{nSab}.

$S_{ab}$ has four simple poles. For $a>b$, these are located at ${\newh\over
i\pi}\th=(a-b),(\newh-a+b),(\newh-a-b)$ and $(a+b)$. Let us  check whether we
can identify these poles as single particle poles. If the pole at
$\th={i\pi\over\newh}(a-b)$ corresponds to a particle in the u-channel then
this
particle has mass
\beq
\mt^2=\mt_a^2+\mt_b^2-2\mt_a\mt_b\cos{\pi\over\newh}(a-b)=
4\mt_n^2\sin^2{(a-b)\pi\over\newh}
\eeq
and this identifies it as particle $\phi_{(a-b)}$. Similarly the pole at
$\th={i\pi\over\newh}(h-a+b)$ is identified as the s-channel pole of the same
particle. In the case $a=b$ the pole
is located at the edge of the physical sheet and does not correspond to a
single particle pole.

The same calculation for the pole at $\th={i\pi\over\newh}(a+b)$ leads to a
mass
\beq\label{massapb}
\mt^2=\mt_a^2+\mt_b^2+2\mt_a\mt_b\cos{\pi\over\newh}(a+b)=
4\mt_n^2\sin^2{(a+b)\pi\over\newh}
\eeq
If $a+b<{h\over2}$ then this is the mass of particle $\phi_{(a+b)}$. But if
$a+b>{h\over2}$ the pole does not appear at the expected position. The particle
$\phi_{(h-a-b)}$ has a mass
\beq
\mt^2_{(h-a-b)}=4\mt_n^2\sin^2{(h-a-b)\pi\over\newh}
\eeq
which is not equal to \reff{massapb} due to the fact that $h\neq \newh$. As we
will now explain, this displacement of the pole is due to the presence of an
anomalous threshold singularity.

In perturbation theory (or in a pure analyticity-unitarity S-matrix approach
\cite{4p}) using renormalized masses, the amplitude $S_{ab}$
for $a+b >n$ has not only a simple pole corresponding to the particle
$\phi_{h-a-b}$ but also neighboring poles produced as anomalous threshold
singularities from the various diagrams in Fig. (1.b,c,d,e,f). Indeed,
using the value of the renormalized masses it is straightforward to
check by means of a dual diagram analysis that the triangle diagrams
produce  pole singularities located at
$\th =\frac{i\pi}{\newh}(a+b)
$
and the crossed box in Fig. (1.d) has a double pole at the same position.
Using standard formulas the coefficients of these poles are
\EQ
T_{ab} ={\frac18} \frac{\sin\frac{\pi}{\newh}(a+b)}{\cos
\frac{\pi}{\newh}a \cos \frac{\pi}{\newh}b}
{}~~~~~~~~R_{ab} = 8 T^2_{ab} \sin \frac{2\pi}{\newh}(a+b)
\EN
respectively, multiplied by appropriate coupling constants.
Specifically, denoting by $\d_{ab}
$ the shift of the S-matrix pole from its
expected position, i.e.
\EQ
\d_{ab} = 4\widetilde{m}^2_n \left( \sin^2
\frac{\pi}{\newh}(a+b) -\sin^2 \frac{\pi}{\newh}(h-a-b) \right)
\EN
and
$
\s =s-\widetilde{m}^2_{h-a-b}
$
we have the contributions from the six diagrams
\bea
(a) &:& ~~~\frac{1}{\s} c^2_{ab,h-a-b} \nonumber\\
(b) &:& ~~~\frac{T_{ab}}{\s
-\d_{ab}}c_{ann}c_{bnn}c_{abnn} \nonumber\\
(c) &:&~~~\frac{T_{ab}}{\s (\s -\d_{ab})}c_{ann}c_{bnn}c_{ab,h-a-b}c_{nn,h-a-b}
\nonumber\\
(d) &:& ~~~\frac{R_{ab}}{(\s - \d_{ab})^2} c^2_{ann}c^2_{bnn} \nonumber\\
(e) &:& ~~~\frac{T^2_{ab}}{(\s -\d_{ab})^2} c^2_{ann}c^2_{bnn}c_{nnnn}
\nonumber\\
(f) &:& ~~~\frac{T^2_{ab}}{\s (\s -\d_{ab})^2} c^2_{ann}c^2_{bnn}
c^2_{nn,h-a-b}
\eea
Here the coupling constants can be obtained directly from the lagrangian
in a lowest order computation, but for a complete comparison higher-order
corrections should be included, as well as contributions from the subleading
parts of the triangle and crossed box diagrams. %\bea
%(a) &:& \frac{1}{\s} ~~~~~~~~~~ (b) ~:~ \frac{T}{\s - \d} ~~~~~
%(c)~:~ \frac{T}{\s (\s - \d )} \nonumber\\
%(d) &:& \frac{R}{(\s -\d )^2} ~~~~~ (e)~:~ \frac{T^2}{(\s - \d )^2}
%~~~~~(f)~:~ \frac{T^2}{\s (\s - \d )^2}
%\ena
%all multiplied by suitable coupling constants.

We have checked that in
the sum of the above terms  the pole at
$\s =0$ cancels, leaving a  simple pole at $\s =\d$ with the correct
$O(\b^2)$ residue
 and indeed reproducing the result obtained from the exact S-matrix.
This explains the shift
of the simple pole in the S-matrix from the expected location
at the (renormalized) mass. We emphasize that the cancellation of the pole at
$\s=0$ is due to a subtle interplay between the location of the anomalous
threshold poles and their residues.

It is interesting to contrast the above discussion of the anomalous poles with
that in the $A^{(2)}(0,2n-1)$ theory \cite{9} which differs from our theory by
the addition of one fermion: there, the bosonic field $\phi_n$ gives rise to
the
same anomalous threshold structure, but this is precisely cancelled by an
identical structure from the fermion, so that the simple particle pole ends up
in the position predicted by the mass formula (of the classical theory).

As an  aid to further discussion we note that many of the results and
expressions derived in Ref. \cite{9} for  some of the amplitudes
and the dual diagram constructions in
the $A^{(2)}(0,2n-1)$ theory can be taken over to our theory. Also,
the exact S-matrix
elements  $S_{an}$ and $S_{ab}$ have  the same form  in the two theories,
provided the blocks are reinterpreted  according to our definitions above
i.e. with $h \rightarrow \newh$.

We consider now the double poles in the amplitude $S_{an}$ which occur at
\EQ
\th =\frac{i\pi}{\newh}\left(\frac{\newh}{2} +2p-a\right)~~~~~~~ p=1,2,...a-1
\EN
It is straightforward to check that indeed these can be accounted for by
both uncrossed and crossed box diagrams with dual diagrams similar to the
ones of Fig. 12 in Ref. \cite{9}. We emphasize that although
the construction is done with the renormalized masses, so that
the lengths of the lines are  different from what they were in the above
reference, nonetheless the dual diagrams exist for both the crossed and
the uncrossed boxes. The only change is in the location of the double
poles, and this agrees with their location in the exact S-matrix.
 To one-loop order the
coefficients of these double poles are the same as for the superalgebra
case (where these contributions are also from bosons only),
 and  since our $S_{an}$ has the same structure there is no
need for any further checks at the one-loop level.

A new feature appears when we examine the double poles of the $S_{ab}$
amplitudes. For example, in the superalgebra case one obtained double
poles from both an uncrossed and a
crossed box, cf. Fig. 13 of reference \cite{9} for the case of the
$S_{n-1,n-1}$
amplitude. Now however, using the
renormalized masses, the dual diagram for the uncrossed box can no
longer be drawn in the plane and it would seem that although a double
pole is still produced by the crossed box its coefficient would not
match that of the exact S-matrix; the contribution from the uncrossed box
is needed. The explanation is provided by realizing that in addition to
the box diagram there are seemingly higher-order diagrams where one of the
internal lines is replaced by the set of diagrams appearing in Fig. 1
which are responsible for the displacement of the single particle pole. In
going
through the Landau-Cutkoski analysis for the location and nature of
singularities, one realizes then that a double pole is indeed produced, with
the
correct residue. Equivalently one can perform  the dual diagram analysis using
not the actual particle mass $\widetilde{m}^2_{h-a-b}$ but that
corresponding to the actual pole of the S-matrix,
namely $\widetilde{m}^2_{h-a-b} + \d_{ab}$. In this manner all the double poles
of the exact S-matrix can be accounted for, and aside from verifying that the
higher-order coefficients are correctly given we claim to have checked the
self-consistency of the S-matrix we have constructed.

To make contact with the lagrangian of the $a_{2n-1}^{(2)}$ theory
we have performed  some further perturbation theory checks.
We assert that our S-matrix agrees at tree level
with that computed from the lagrangian in \reff{action}. This is obvious for
the amplitudes $S_{an}$ and $S_{ab}$ provided we choose
$
\e(\b )= -\frac{\b^2}{4\pi h} +O(\b^4)
$,
since the corresponding amplitudes of the $A^{(2)}(0,2n-1)$ theory were
checked to agree with the tree level amplitudes and the bosonic
lagrangian of that theory is identical to our lagrangian. On the other
hand our $S_{nn}$ amplitude has a rather different form, and a separate
check was necessary. In this respect, extending the product in \reff{Snn} to
include the $a=0$ term was crucial.

We observe that our S-matrix predicts a very specific
manner in which the masses of the theory renormalize:
\EQ
\widetilde{m}_a^2 = 4 \widetilde{m}_n^2 sin^2\left( \frac{\pi a}{h}(1+\e)
\right) \EN
This is consistent with the one-loop mass corrections in \reff{dm}, and
already provides a nontrivial check of our S-matrix and of the
restrictions that follow from the bootstrap. Of course it would be
interesting to perform a direct two-loop calculation of the mass shifts
and compare with those predicted here.

As already discussed, there is agreement at the one-loop level
between the coefficients of the double poles in the exact S-matrix and
those computed perturbatively. However, one-loop checks also allow us
to determine the coupling constant dependence of the S-matrix.
We have calculated from the lagrangian the one-loop corrections to
the three-point functions for the $a_3^{(2)}$ theory. These corrections
arise from three sources: genuine vertex corrections, wave-function
renormalization effects, and a rescaling of the mass scale in the lagrangian
(which rescales all the couplings) that we performed in order to keep the mass
of the particle $\phi_n$ at its classical value. The result, when substituted
in the perturbative S-matrix, can be compared with that from the
exact S-matrix through order $\b^4$ and $\e^2$ respectively. We find agreement
provided we choose
\EQ
\e = -\frac{\b^2}{4\pi h} +(h+1)\left(\frac{\b^2}{4\pi h}\right)^2 +O(\b^6)
\EN
and therefore, to this order
\EQ
B\equiv -\frac{2h\e}{1+\e} = \frac{1}{2\pi}\frac{\b^2}{1+\frac{\b^2}{4\pi}}
\EN
which is the standard dependence on the coupling constant that one has
found in simply-laced theories \cite{10}.

Finally, as a further perturbative check, we have computed for the $a_3^{(2)}$
 case,
up to
one-loop order, the spin 3 charges of the two particles of the theory,
and found agreement with those predicted by the exact S-matrix \cite{6}.

Let us summarize: we  have used very little information from the lagrangian
of the $a_{2n-1}^{(2)}$ Toda theory to determine the S-matrix in
eqs.(\ref{nSnn},\ref{nSan},\ref{nSab}). We assumed the existence
of a higher spin symmetry but did not need its explicit form. We admitted the
fact that the classical masses renormalize in a nontrivial manner
but did not make any assumptions about the form
of this renormalization; and we used the existence of
three-point couplings $c_{nna}$ but did not use their values. This
information was almost sufficient to determine the S-matrix using unitarity,
real analyticity and the bootstrap
principle. We chose not to include more CDD poles
and zeros in
$S_{nn}$ and we chose the smallest number of poles in $S_{an}$.
We obtained an S-matrix which satisfies all the bootstrap consistency
conditions except those arising from certain displaced poles in $S_{ab}$.
We explained the displaced simple poles
and all the  higher-order poles as arising from anomalous threshold
singularities. Finally, by comparing with perturbative calculations, we
demonstrated that we are indeed dealing with the S-matrix of the
$a_{2n-1}^{(2)}$ Toda theory. In particular, we obtained the relation between
the parameter $B$ and the coupling constant $\b$ of the Toda theory. Obviously,
given the present state of the technology, this could only be done to low
orders of perturbation theory, but in this respect the situation is
similar to that of most simply-laced Toda theories.

We comment on two other features: first, application of the
thermodynamic Bethe Ansatz \cite{11}  leads to the prediction of the central
charge in the conformal limit, $c=n$, as one might expect. Second,
it does not appear that a minimal S-matrix exists for this theory.
One cannot drop the blocks involving the dependence on the coupling
constant through the parameter $B$ since this would lead to a larger number of
poles in \reff{Sanlong} than can be accounted for.

Finally, it appears that with a minor modification our S-matrix also
describes scattering for the Toda theory based on the Lie superalgebra
$A^{(4)}(0,2n-1)$. We recall \cite{4} that this theory can be
obtained from the $A^{(2)}(0,2n-1)$ theory by dropping the boson $\phi_n$
instead of the fermion $\psi$. Consequently the particle masses are
again shifted away from their classical values, this time by an amount
which is just the negative of the one in \reff{dm}. Defining for $S_{\psi
\psi}$
the same expression as for $S_{nn}$ but choosing $\eta_0=0$ in \reff{choice}
and replacing $\e$ by $-\e$ leads by bootstrap to a $S_{\psi a}$ and $S_{ab}$
which have the same form as the previous $S_{na}$ and $S_{ab}$
and provide a consistent description of the S-matrix for this
Toda theory.

Details of our work and extension to other nonsimply-laced theories
will be presented in  separate publications \cite{6,7}.

\vskip 8 cm
\begin{figure}
\special{dvitops: import deliu.eps \the\textwidth 12cm}
\caption{Contributions to the $S_{ab}$ amplitude which are responsible for the
shift in the particle pole position.}
\end{figure}

\end{document}